*Giannis Haralabopoulos, Ph.D. Computer Science*

(The following plan is a short summary, of a more extensive research, included in my Ph.D. dissertation)

# An Anonymous Online Social Network of Opinions

## Abstract

Research interest on Online Social Networks (OSNs), has increased dramatically over the last decade, mainly because online networks provide a vast source of social information. Graph structure, user connections, growth, information exposure and diffusion, are some of the most frequently researched subjects. However, some areas of these networks, such as anonymity, equality and bias are overlooked or even unconsidered. In the related bibliography, such features seem to be influential to social interactions. Based on these studies, we aim at determining how universal anonymity affects bias, user equality, information propagation, sharing and exposure, connection establishment, as well as network structure. Thus, we propose a new Anonymous Online Social Network, which will facilitate a variety of monitoring and data analysis.

## Literature

In [1], the authors point out that personal anonymity within 4chan, shapes a strong communal identity among the great number of individuals. The author of [2], notes that online anonymity assists users to overcome potential boundaries of identity or culture, thus promoting an open and free exchange of ideas, while in [3] the authors analyze the effects of anonymity in social restaurant reviewing sites. They found out that while anonymity does not influence rating policies, the lack of a profile negatively affects ratings themselves.

Several social characteristics, such as tightly knit communities, two step flow of communication [4] and power law distributions [5], have been universally linked with Online Social Networks. Barabási and Oltvai, in [6], observed that when new users join a network, they tend to connect to an already strong node of the network, thus creating non-random connectivity power law distributions, which are evident in every Online Social Network, such as Facebook [7], Twitter[1] and Reddit[2]. While in [8], authors note that densely knit communities and the patterns of interactions the communities of a greater network have important implications to the functioning of the network.

Furthermore, bias in social networks, are insufficiently investigated. Author of [9] argues that users in Twitter are biased positively towards their friends and negatively to unknown users. The researcher also notes that the name of a user, affects our perception in respect to the information the user shares. A very interesting work on Persuasion Bias [10] describes how individuals and groups are influenced. The influence on group opinions doesn't only depend on accuracy, but also on how well connected one is the user on the given communication network.

---

[1] http://krugman.blogs.nytimes.com/2012/02/08/the-power-law-of-twitter/
[1] http://blog.luminoso.com/2012/02/09/twitter-followers-do-not-obey-a-power-law-or-paul-krugman-is-wrong/
[2] http://www.reddit.com/r/TheoryOfReddit/comments/2lp2i6/a_ranksize_distribution_for_the_top_500/

**Aim**

Our aim is three-fold: i) to study the effects of anonymity in Online Social Networks and ii) further analyze equality and bias. In one hand, it is apparent that these goals cannot be completely achieved in any form of established OSN[3,4,5], because a public profile creation is required and data access is restricted in all of them. On the other hand, anonymous online networks[6,7] or applications[8] that do not require a public profile, are either focused in different forms of communication, or their data collection is insufficient. Thus, our third goal is to provide unrestricted data access to interested researchers and individuals, with outmost respect to user privacy.

In order to achieve these 3 goals, we propose the creation of a new online social network. A network that will protect user's privacy, have an open data structure, try to eliminate relationship bias and promote equality. As a feasible method to reach the simultaneous success of our goals, we choose universal anonymity. However, as an OSN characteristic, anonymity needs a certain degree of extensive research in a monitored environment, before concluding to its catalytic factor in user interaction. Our proposed network, will also serve as an extensive experimental platform for anonymity and/or similar concepts.

**Project**

Our project's schedule will be spilt into three parts and will be based on the needs of each Work Package (WP). Ideally, our network should reach a -statistically- significant size after a number of months.

Work Package 1: Technical foundations. WP1 will develop the platform of our proposed OSN with a strong underlying database structure, which will allow monitoring of user relations and content linkage. Several technical issues will be addressed, such as post topics, profiles, search and keywords functions, content sorting and recommendations algorithms.

Work Package 2: Testing. WP2 will include both the testing and tweaking of the platform. Content, database, presentation and more will be addressed as the network is populated. With its expansion, users and their relations will evolve, data volume will increase and methods will be reassessed.

---

[3] https://developers.facebook.com/docs
[4] https://dev.twitter.com/rest/public/rate-limiting
[5] https://developers.google.com/+/api/
[6] http://www.4chan.org/
[7] https://whisper.sh/
[8] http://www.yikyakapp.com/

Work Package 3: Experimentation. In this WP we will perform a multitude of social network experiments, ranging from multiple OSN simultaneous posting, to analyzing the effects of simple structural changes. Our aim is to create an open form of anonymous online social network, which will create new research opportunities for any interested individual.

## Discussion

We look forward into creating such a free and open social network, from a researcher's standpoint. Our idea is constantly evolving, through dialogue and opinion exchanges. The diverse and interdisciplinary character of a renowned institution, would only improve the foundations of our concept. Even if our proposed network will not reach a wide adoption, its unique character will surely provide a new research platform for computer, information and social scientists.